\documentclass{emulateapj}
\usepackage{times}
\newcommand\bb[1] {\mbox{\boldmath{$#1$}}}

\begin{document}

\shorttitle{\textsc{Scaling Laws in MRI-driven Turbulence}}
\shortauthors{\textsc{PESSAH, CHAN, \& PSALTIS}}
 
\title{\textsc{Angular Momentum Transport in Accretion Disks: \\
Scaling Laws in MRI-driven Turbulence}}

\author{Martin E. Pessah$^{1,3,4}$, Chi-kwan Chan$^{2,4}$, and Dimitrios Psaltis$^{4,3}$}
\altaffiltext{1}{Institute for Advanced Study, Princeton, NJ, 08540}
\altaffiltext{2}{Institute for Theory and Computation, Harvard-Smithsonian Center for Astrophysics, Cambridge, MA, 02138}
\altaffiltext{3}{Astronomy Department, University of Arizona, Tucson, AZ, 85721}
\altaffiltext{4}{Physics Department, University of Arizona, Tucson, AZ,
  85721}

\email{mpessah@ias.edu (MEP)}

\begin{abstract}
  We present a scaling law that predicts the values of the stresses
  obtained in numerical simulations of saturated MRI-driven turbulence
  in non-stratified shearing boxes. It relates the turbulent stresses
  to the strength of the vertical magnetic field, the sound speed, the
  vertical size of the box, and the numerical resolution and predicts
  accurately the results of 35 numerical simulations performed for a
  wide variety of physical conditions.  We use our result to show that
  the saturated stresses in simulations with zero net magnetic flux
  depend linearly on the numerical resolution and would become
  negligible if the resolution were set equal to the natural
  dissipation scale in astrophysical disks.  We conclude that, in
  order for MRI-driven turbulent angular momentum transport to be able
  to account for the large value of the effective alpha viscosity
  inferred observationally, the disk must be threaded by a significant
  vertical magnetic field and the turbulent magnetic energy must be in
  near equipartition with the thermal energy. This result has
  important implications for the spectra of accretion disks and their
  stability.
\end{abstract}


\section{Introduction}

One of the key unsolved problems in accretion disk physics is the
precise nature of the mechanism that allows for angular momentum to be
transported outwards in the disk. In the standard theory, ones copes
with this problem by arguing that the stress responsible for angular
momentum transport, due to turbulence and magnetic fields, is
proportional to the local pressure, i.e., $\bar{T}_{r\phi} \equiv q
\alpha \bar{P}$. Here, $\alpha$ is a dimensionless constant of the
order of, but smaller than, unity and $q\equiv - d\!\ln \Omega/d\!\ln
r$ characterizes the local shear \citep{SS73, KFM98, FKR02}.

Despite the fact that the standard parametrization leads to a disk
model in which the energy generation rate is determined mostly by
energy balance and depends weakly on the adopted prescription
\citep{BP99}, almost every aspect of the disk structure depends
explicitly on this assumption~\citep{KFM98, FKR02}.  Therefore
calculating the value of $\alpha$, or more generally, testing whether
the assumed relationship between stress and pressure is adequate, is
of fundamental importance.  These questions, however, lie outside the
scope of the standard theory, making necessary the identification of a
specific physical mechanism for angular momentum transport.

Over the last decade, magnetohydrodynamic (MHD) turbulence driven by
the magnetorotational instability (MRI; Balbus \& Hawley 1991; 1998)
has emerged as the most promising candidate to enable angular momentum
transport in astrophysical disks. The development of three-dimensional
MHD numerical codes has led to a detailed study and characterization
of angular momentum transport in turbulent magnetized disks. This has
motivated numerical estimations of the $\alpha$ parameter and, more
generally, the search for saturation predictors to describe the
turbulent state \citep[see, e.g.,][]{HGB95, HGB96, Brandenburg95,
  Gammie98, PCP07}.

It has long been recognized that the $\alpha$ parameter is not a
constant. It is known to depend, among other things, on the strength
and geometry of the magnetic field and, perhaps more uncomfortably, on
the size of the simulation domain and the resolution. In spite of
this, it has not been possible to disentangle numerical from physical
dependencies in a clear way \cite[see, e.g.,][]{HGB95, HGB96,
  Brandenburg98, Sanoetal98, Sanoetal04}.  Being able to distinguish
between these two types of dependencies is vital if we seek to use the
results of numerical simulations to build angular momentum transport
models, and eventually global disk models, beyond the standard
prescription \citep[see, e.g.,][]{KY95, Ogilvie03, PCP06b, PCP07}.

The search for a mechanism for the saturation of the MRI-driven
turbulence has been a long-sought for goal since the appreciation of
the relevance of the MRI to astrophysical disks. By necessity, a key
piece of this puzzle consists of understanding the role that the
various factors, both physical and numerical, play in the saturated
state.  In this \emph{Letter}, we provide an expression that describes
the transport of angular momentum in shearing MHD boxes, based on a
series of local numerical simulations carried out by
\cite{Sanoetal04}. In particular, we are able to disentangle how the
characteristics of the MHD flow depend on the various physical
(pressure, magnetic field, etc.) and numerical (box size and
resolution) factors.

\section{Characteristic Scales in Numerical Simulations}

Numerical studies addressing the local dynamics of three-dimensional,
differentially rotating, turbulent magnetized flows are often carried
out in the shearing box approximation.  This consists of a first order
expansion in the variable $r-r_0$ of all the quantities characterizing
the flow at the fiducial radius $r_0$.  The goal of this approach is
to retain the most relevant physics governing the dynamics of the MHD
fluid in a locally-Cartesian coordinate system co-orbiting and
corotating with the background flow with local velocity $\bb{v}_0 =
r_0\,\Omega_0 \check{\bb{\phi}}$.
  
In the shearing box framework, all the physical variables are usually
normalized using the initial density, $\rho_0$, and the angular
velocity, $\Omega_0$.  For a more detailed discussion concerning the
physical approximations and numerical implementations involved in the
shearing box approach see \cite{HGB95}.  In an unstratified shearing
box there are three possible scales of length relevant to the vertical
direction: {\em (i)\/} The first scale is the size of the box $L$.
{\em (ii)\/} The second scale is the wavelength corresponding to the
most unstable MRI mode, i.e., $\lambda_{\rm MRI} \equiv 2\pi
\sqrt{16/15}\, \bar{v}_{{\rm A}z}/\Omega_0$, where $\bar{v}_{{\rm
    A}z}$ is the average Alfv\'en velocity associated with the
vertical component of the magnetic field, ${v}_{{\rm A}z} \equiv
B_{z}/(4\pi \rho_0)^{1/2}$, and we have assumed a Keplerian shear
profile (i.e., $q=3/2$). {\em (iii)\/} The last scale is associated
with the sound speed: $H \equiv (2/\gamma)^{1/2}c_{s}/\Omega_0$, where
$\gamma$ is the ratio of specific heats and $c_{\rm s}$ is the speed
of sound. In a stratified disk, $H$ is the pressure scale-height and,
for simplicity, we will refer to it as such hereafter, even though the
particular simulations we will be discussing are not stratified.

Most systematic studies carried out to characterize the turbulent
state driven by the MRI consider non-radiative flows in which the
internal energy grows in time due to magnetic dissipation.
Furthermore, most of the initial efforts to characterize the saturated
state of the MRI employed numerical schemes that evolved the internal
energy, as opposed to the total energy. In these cases the reported
values of $\alpha$, as well as the expressions for different predictor
functions, where obtained considering the initial pressure $P_0$.  In
such simulations, however, gas pressure increases linearly with time
and this effect must be considered in order to understand its effects
on the stresses responsible for angular momentum transport.

\cite{Sanoetal04}, carried out a series of numerical simulations in
order to investigate the effects of the evolving pressure on the
turbulent state.  They employed an algorithm that solves the energy
equation in terms of the total energy, which allowed them to keep
better track of the energy budget of the flow. By considering the
vertical extent of the box $L$ as the unit of length, they were able
to examine the dependence of the efficiency of angular momentum
transport on the gas pressure. In this case, both the gas pressure
$P_0$ and the mean Alfv\'en velocity $\bar{v}_{{\rm A}z}$, are
independent parameters and determine the ratios $H_0/L$ and
$\lambda_{\rm MRI}/L$, respectively, where the subscript zero
indicates initial values.

\section{Scaling Laws in MRI-driven Turbulence} 
\label{sec:scaling_laws}

\cite{Sanoetal04} performed an extensive set of simulations with zero
net magnetic flux through the domain, initialized with a vertical
magnetic field $B_{0z}(r) = B_{0z}\sin[2\pi (r-r_0)/L_r]$, and runs
with a uniform magnetic field, $B_{0z}$, perpendicular to the disk
midplane\footnote{Note that in both of these cases the magnetic flux
  through the domain is conserved because of the adopted vertical
  periodic boundary conditions.}. They investigated adiabatic
($\gamma=5/3$) as well as isothermal ($\gamma=1.001$) equations of
state. They also performed a suite of runs to study the effects of
Ohmic dissipation but we will not discuss them here. For all their
models they used a shearing box with dimensions $L=L_r=L_z=1$ and
$L_\phi = 4L$, and a grid of $32\times128\times32$ zones. The scales
of density and time are the same as in \cite{HGB95}, i.e., $\rho_0=1$
and $\Omega_0=10^{-3}$.


\begin{figure}
  \includegraphics[width=\columnwidth,trim=0 10 0 10]{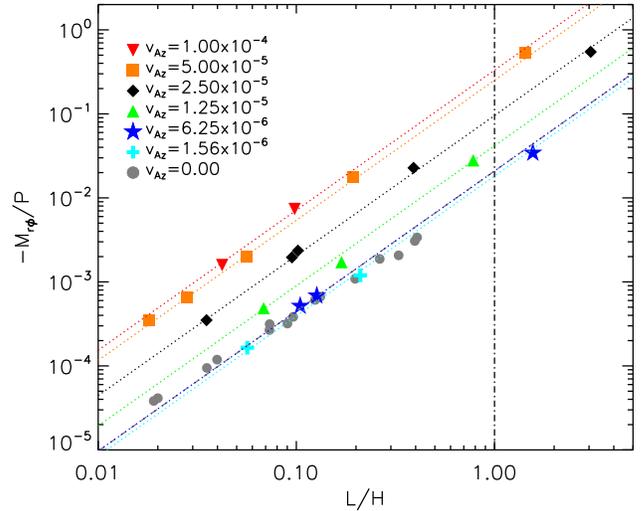}
  \caption{Dimensionless magnetic stress, normalized by the
    instantaneous pressure, as a function of a dimensionless measure
    of the box size with respect to the evolving scale-height, $H=H_0
    (\bar{P}/P_0)^{1/2}$, for the numerical simulations of
    \cite{Sanoetal04}. The various types of symbols label simulations
    according to the value of the ratio $\bar{v}_{{\rm
        A}z}/L\Omega_0$, with $L=1$ and $\Omega_0=10^{-3}$.  The {\it
      dotted lines} are the best fits for each set of simulations with
    constant ratio $\bar{v}_{{\rm A}z}/L\Omega_0 \ne 0$. The {\it
      dashed line} is the best fit for all the simulations with zero
    net magnetic flux. In all the cases a constant slope of 5/3, i.e.,
    $\bar{M}_{r\phi}/\bar{P} \propto (L/H)^{5/3}$, provides a
    remarkably good description of the simulation results.}
\label{fig:fig1}
\end{figure}

Their choice of initial conditions spans six orders of magnitude in
the initial pressure $P_0$ and two orders of magnitude in the Alfv\'en
speed.  Figure~\ref{fig:fig1} shows the dimensionless ratios
$\bar{M}_{r\phi}/\bar{P}$ as a function of the ratio $L/H$
characterizing the turbulent states reached by the adiabatic and
isothermal simulations listed in Tables~1 and~2 in
~\cite{Sanoetal04}. Here, $\bar{M}_{r\phi}\equiv \langle\langle
\delta\!B_r \delta\!B_\phi\rangle\rangle/4\pi$, $\bar{P}$, and $H$
stand for the volume- and time- averaged values of the magnetic
stress, pressure, and, equivalent scale-height $H=H_0
(\bar{P}/P_0)^{1/2}$.

The tight correlations followed by simulations characterized by the
same values of mean Alfv\'en velocities, including the class of
simulations with zero net magnetic flux, i.e., $\bar{v}_{{\rm A}z}=0$,
suggest the scaling
\begin{eqnarray}
\left. \frac{\bar{M}_{r\phi}}{\bar{P}}\right|_{\bar{v}_{{\rm A}z}/L\Omega_0 =
{\rm const.}} \propto \,\left(\frac{L}{H}\right)^{5/3} \,.
\end{eqnarray}

\begin{figure}
  \includegraphics[width=\columnwidth,trim=0 10 0 10]{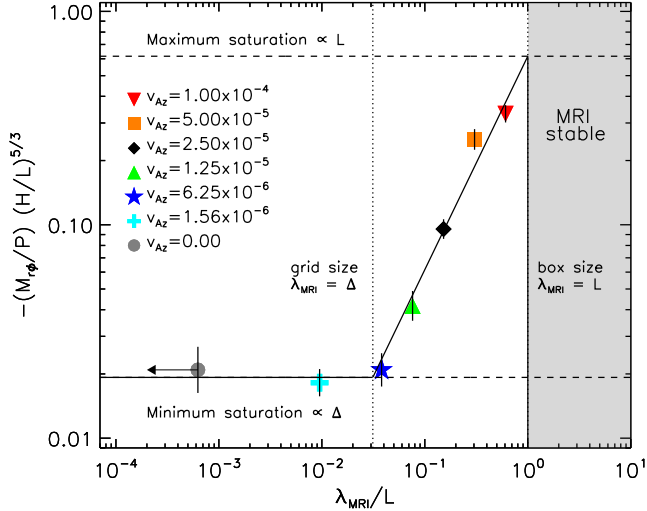}
  \caption{Dimensionless magnetic stress, $-\bar{M}_{r\phi}/\bar{P}$,
    multiplied by $(H/L)^{5/3}$, as a function of the wavelength
    corresponding to the most unstable MRI mode, $\lambda_{\rm
      MRI}$. The various types of symbols correspond to the best fit
    values characterizing each class of simulations according to the
    value of the ratio $\bar{v}_{{\rm A}z}/L\Omega_0$, as inferred
    from Figure~\ref{fig:fig1}.  The simulations with zero mean
    magnetic flux, i.e., $\bar{v}_{{\rm A}z}=0$, are displayed at some
    arbitrary value for visualization purposes only. The error bars
    quantify the scatter within each class of simulations around the
    corresponding mean values. Vertical {\it dotted lines} represent
    the values at which the most unstable MRI wavelength equals the
    grid size and the size of the box, respectively, i.e.,
    $\lambda_{\rm MRI} = \Delta = 1/32$ and $\lambda_{\rm MRI} = L =
    1$. The overall dependence of the saturation level on the ratio
    $\lambda_{\rm MRI}/L$ ({\it solid line}) is given by the
    saturation predictor~(\ref{eq:maxwell_predictor}).  Numerical
    simulations for which $\lambda_{\rm MRI} > L$ are stable to the
    MRI ({\it shaded region}).}
  \label{fig:fig2}
\end{figure}

The best fits to the data are shown in Figure~\ref{fig:fig1} with a
{\it dashed line} and {\it dotted lines} for simulations with zero and
non-zero net magnetic flux, respectively. This scaling\footnote{Note
  that fixing all the slopes to 5/3 in Fig.~\ref{fig:fig1} allows us
  to describe all the simulations on the same footing while leading to
  small residuals in the best fit amplitudes (see
  Fig.~\ref{fig:fig2})} can be used to remove the dependence of the
ratio $\bar{M}_{r\phi}/\bar{P}$ on $L/H$ for each set of simulations
with the same value of the ratio $\bar{v}_{{\rm A}z}/L\Omega_0$.  The
result is shown in Figure~\ref{fig:fig2}, where the various types of
symbols label the best fit values characterizing each class of
simulations according to their value of $\bar{v}_{{\rm
    A}z}/L\Omega_0$.  The error bars quantify the scatter within each
class of simulations from the best fit values obtained in
Figure~\ref{fig:fig1}.  These average values follow a simple, yet
tight, correlation with the associated values of $\lambda_{\rm MRI}$.


The vertical {\it dotted lines} in Figure~\ref{fig:fig2} represent the
values at which the most unstable MRI wavelength equals the size of
the box and the grid size, respectively, i.e., $\lambda_{\rm MRI} = L
= 1$ and $\lambda_{\rm MRI} = \Delta = 1/32$. Note that numerical
simulations for which $\lambda_{\rm MRI} > L$ are stable to the MRI
({\it shaded region}).  The saturation of the simulations for which
$\Delta < \lambda_{\rm MRI} \le L$ is linearly proportional to
$\lambda_{\rm MRI}$, i.e.,
\begin{eqnarray}
\frac{\bar{M}_{r\phi}}{\bar{P}} \left(\frac{H}{L}\right)^{5/3} \propto
\lambda_{\rm MRI} \,.
\end{eqnarray}
Note that \cite{Sanoetal04} concluded that $\bar{M}_{r\phi} \propto
\bar{v}_{{\rm A}z}^{3/2} \propto \lambda_{\rm MRI}^{3/2}$. This
steeper dependence does indeed provide a good description of the
simulations with $\bar{v}_{{\rm A}z} = 5\times10^{-5},
2.5\times10^{-5}, 1.25\times10^{-5}$. However, a linear dependence on
$\bar{v}_{{\rm A}z}$ leads to a better overall description of all the
simulations with $\Delta < \lambda_{\rm MRI} \le L$, i.e., including
those with $\bar{v}_{{\rm A}z} = 6.25\times10^{-6}$ and $\bar{v}_{{\rm
    A}z} = 1\times10^{-4}$.

There is a clear departure from the linear dependence of the saturated
stress on $\bar{v}_{{\rm A}z}$ for the simulations for which
$\lambda_{\rm MRI} \le \Delta$. All of these runs saturate at the same
value of $(\bar{M}_{r\phi}/\bar{P})(H/L)^{5/3}$ regardless of the
value of the magnetic flux through the vertical boundary.  The fact
that this minimum value of the stresses at saturation (lower {\it
  dashed line} in Figure~\ref{fig:fig2}) is equal to 1/32 of the
maximum possible value, corresponding to $\lambda_{\rm MRI} = L$
(upper {\it dashed line} in the same Figure), strongly suggests that
this floor is entirely set by the grid size.  Extrapolating this
behavior to lower values of the grid scale suggests that this minimum
saturation level is itself linearly proportional to the size of the
grid\footnote{After this paper appeared on the preprint server,
  Fromang and Papaloizou posted a paper (arXiv:0705.3621) in which
  they perform ideal MHD simulations for zero-net-field shearing boxes
  with increasing resolution. They found that the saturated stresses
  depend linearly on resolution, in agreement with the results
  presented here.}. This is consistent with the results found in
\cite{HGB96}; see in particular their Figure 8, where the final plasma
$\beta$'s for simulations with zero net magnetic flux and uniform
vertical magnetic fields is shown.  The ratio of the final plasma
$\beta$ for the simulation with highest uniform field to the average
final plasma $\beta$ corresponding to the zero net magnetic flux runs
is indeed close to $\Delta =1/32$.

The overall dependence of the saturation level on the ratio
$\lambda_{\rm MRI}/L$, i.e., the {\it solid line} in
Figure~\ref{fig:fig2}, is described by the function
\begin{eqnarray}
\label{eq:maxwell_predictor}
\frac{\bar{M}_{r\phi}}{\bar{P}} \simeq \, -0.61
\left(\frac{L}{H}\right)^{5/3} \!\!\!\!\!  \times \left\{
\begin{array}{ccr} 
\Delta/L            & \textrm{ if } & \lambda_{\rm MRI} \le  \Delta \\
\lambda_{\rm MRI}/L & \textrm{ if } & \Delta < \lambda_{\rm MRI} \le L \\
0                   & \textrm{ if } & \lambda_{\rm MRI}  > L
\end{array} 
\right. . \nonumber \\ 
\end{eqnarray}
This saturation predictor is consistent, in the intermediate regime,
with the one obtained by \cite{HGB95} that we also used in an earlier
work (Pessah et al.\ 2006b).  Indeed, substituting in this expression
$\bar{P}=P_0=\rho_0 (H_0 \Omega_0)^2$, $H=H_0=L$, and considering that
the magnetic stress is roughly half of the magnetic energy in the
fluctuations, $\bar{M} \equiv
\langle\langle\delta\!B^2\rangle\rangle/8\pi$ \citep[][and references
therein]{BPV06}, leads to
\begin{eqnarray}
\bar{M} = 1.2 \rho_0 \,(L\Omega_0)\,(\lambda_{\rm MRI}\Omega_0)  \,,
\end{eqnarray}
which is identical to equation (18) of~\cite{HGB95}.

The $5/3$ scaling in the saturation
predictor~(\ref{eq:maxwell_predictor}) is reminiscent of the
Kolmogorov spectrum of turbulence. The prominent role of $\lambda_{\rm
  MRI}$ in this expression is another indication that the MRI
continues to pump the turbulence, even in the non-linear state (see
also Pessah et al.\ 2006a, 2006b). The dependence of the stress on the
vertical size of the box, $L$, must be an artifact of the periodic
boundary conditions, which do not allow for any turbulent energy to
escape the domain of solution. The dependence on the resolution must
be related to numerical dissipation at the grid scale, and might,
therefore, change if a different MHD algorithm were used for the
simulations. Finally, the role of the ``scale-height'', $H$, is very
difficult to understand, since the simulations are not stratified and
this length scale is too large to be resolved. It may arise from a
non-linear coupling between sound waves and MHD modes or it might
simply be related to numerical dissipation through the dependence of
the Courant time step on sound speed.

\section{Implications and Discussion}

In shearing-box simulations of Keplerian flows, the Maxwell and
Reynolds, $\bar{R}_{r\phi}\equiv\langle\langle \rho \delta\!v_r
\delta\!v_\phi\rangle\rangle$, stresses follow a tight correlation
\citep[see][and references therein]{PCP06a} with
$-\bar{M}_{r\phi}/\bar{R}_{r\phi} \simeq 4$. Taking this into account,
we can write for the total stress and the effective alpha viscosity
\begin{eqnarray}
\label{eq:stress}
\frac{\bar{T}_{r\phi}}{\bar{P}} = q\alpha \simeq 
\, 0.75 \left(\frac{L}{H}\right)^{5/3} \!\!\!\!\!  \times \left\{
\begin{array}{ccr} 
\Delta/L            & \textrm{ if } & \lambda_{\rm MRI} \le  \Delta \\
\lambda_{\rm MRI}/L & \textrm{ if } & \Delta < \lambda_{\rm MRI} \le L \\
0                   & \textrm{ if } & \lambda_{\rm MRI}  > L
\end{array} 
\right. , \nonumber \\ 
\end{eqnarray}
where $q=3/2$ for a Keplerian disk.  This is a remarkable result. This
expression accurately describes the overall dependence of the
saturated state for 35 numerical simulations spanning six orders of
magnitude in initial pressure, encompassing domains with zero and
non-zero net magnetic flux, as well as adiabatic and isothermal
equations of state.

Note that the very small values of the effective alpha viscosity
reported in the past, which are unable to account for observations of
astrophysical disks~\citep{KPL07}, correspond to shearing box
simulations with vertical extents that are small compared to the
equivalent pressure scale-height.  One would expect that a realistic
value of the stress is achieved in simulations with $L\simeq
H$. Nevertheless, even having taken this into account, large values of
the alpha viscosity can be achieved only for particular
configurations. This has three important implications for accretion
disk models in which the angular momentum transport is mediated by
MRI-driven turbulence.

First, the saturated stresses in simulations with zero net magnetic
flux are linearly proportional to the numerical resolution. This
implies that if we were able to set the numerical resolution to the
natural dissipation scale in the problem, which is many orders of
magnitude smaller than the pressure scale height, the MRI would be
unable to sustain the necessary turbulent stresses in a Keplerian
shearing box, unless there were a significant magnetic flux through
the vertical box boundaries.

Second, in order for MRI-driven turbulence to account for the large
values of the effective alpha viscosity inferred from observations
($\alpha\gtrsim 0.1$; see King et al.\ 2007), the vertical magnetic
field must grow to a strength such that the most unstable MRI-modes
have wavelengths comparable to the disk scale-height, i.e.,
$\lambda_{\rm MRI}\simeq L\simeq H$. Such vertical fields are only a
small fraction, roughly a few hundredths, of the associated
equipartition field and pose no significant problem to the energy
budget of the accretion flow.  Indeed, the origin of such a small mean
magnetic field perpendicular to the disk midplane need not be external
to the disk. MRI-driven fluctuations can easily give rise to magnetic
fields of this order, perhaps through the combined effects of
shearing, MRI, and Parker instability, as in the mechanism proposed
by~\citet{TP92}. Note that disk stratification is likely to play an
important role in driving helical turbulence and thus in enabling the
development of a global, large-scale magnetic flux
\citep{Brandenburg95,TB04,BT04}.

Finally, in MRI-driven turbulence, the turbulent magnetic energy is
comparable to $\bar{T}_{r\phi}$ \citep{PCP06a} and, hence, will have
to be also comparable to the thermal energy, if $\alpha$ is of order
unity. This implies that the vertical scale height of an accretion
disk is set by both the magnetic and the thermal pressures, with
important implication for the spectrum emerging from each disk annulus
\citep{Blaesetal06} as well as for the viscous and thermal stability
of the disk.

In closing, it is important to remark that the saturation
predictor~(\ref{eq:stress}) may be specific to shearing-box
simulations with periodic boundary conditions and not necessarily
applicable to the local saturation of stresses in stratified
\citep[e.g.,][]{Brandenburg95,MS00} or global simulations
\citep[e.g.,][]{Armitage98, H00, H01}.  In these cases, the physical
mechanism that limits the growth of turbulent magnetic energy maybe
related to magnetic buoyancy or to large meridional circulation. In
both mechanisms, magnetic energy is lost at large scales and,
therefore, the dependence of the saturated stress on the numerical
resolution and the sound speed may disappear. Understanding the
physical origin of the saturation predictor~(\ref{eq:stress}) and
comparing it to simulations of stratified shearing boxes will resolve
these issues.

\acknowledgments{We thank Eric Blackman, Jeremy Goodman, Gordon
  Ogilvie, and Jim Stone for valuable comments and discussions.}




\begin{thebibliography}{}

\bibitem[Armitage(1998)]{Armitage98}
{Armitage, P.\ J. 1998, ApJ, 501, L189}

\bibitem[Balbus \& Hawley(1991)]{BH91}
{Balbus, S.\ A. \& Hawley J.\ F. 1991, ApJ, 376, 214}

\bibitem[Balbus \& Hawley(1998)]{BH98}
{---------. 1998, Rev. Mod. Phys., 70, 1}

\bibitem[Balbus \& Papaloizou(1999)]{BP99}
{Balbus, S.\ A. \& Papaloizou J.\ C.\ B. 1999, ApJ, 521, 650}

\bibitem[Blackman, Penna, \& Varniere(2006)]{BPV06}
{Blackman, E.\ G., Penna, R.\ F., \& Varniere, P. 2006, [astro-ph/0607119]}

\bibitem[Blackman \& Tan(2004)]{BT04}
{Blackman, E.\ G. \& Tan, J.\ C. 2004, Ap\&SS, 292, 395}

\bibitem[Blaes {et~al.}(2006)Blaes, Davis, Hirose, Krolik, \& Stone]{Blaesetal06}
{Blaes, O.\ M., Davis, S.\ W., Hirose, S., Krolik, J.\ H., \& Stone, J.\ M. 2006, ApJ, 645, 1402}

\bibitem[Brandenburg, Nordlund, Stein, \& Torkelsson(1995)]{Brandenburg95} 
{Brandenburg, A., Nordlund, A., Stein, R.\ F., \& Torkelsson, U. 1995, ApJ, 446, 741}

\bibitem[Brandenburg(1998)]{Brandenburg98}
  {Brandenburg, A. 1998, in Theory of Black Hole Accretion Discs,
    Abramowicz, M.\ A., Bj\"ornsson, G., \& Pringle, J.\ E., eds., Cambridge
    University Press (Cambridge)}

\bibitem[Frank, King, \& Raine(2002)]{FKR02}
{Frank, J., King A., \& Raine, D.\ J. 2002, Accretion Power in
Astrophysics, 3rd edn., Cambridge University Press (Cambridge)}

\bibitem[Fromang \& Papaloizou(2007)]{FP07}
{Fromang, S. \& Papaloizou, J. 2007, [arXiv:0705.3621]}

\bibitem[Gammie(1998)]{Gammie98}
{Gammie C.F. 1998, in Accretion Processes in Astrophysical Systems: 
Some Like it Hot!, S.S., Holt, T.R. Kallman, AIPC 431, 99}

\bibitem[Hawley(2000)]{H00}
{Hawley, J.\ F. 2000, ApJ, 528, 462}

\bibitem[Hawley(2001)]{H01}
{---------. 2001, ApJ, 554, 534}

\bibitem[Hawley {et~al.}(1995)Hawley, Gammie, \& Balbus]{HGB95}
{Hawley, J.\ F., Gammie, C.\ F., \& Balbus, S.\ A. 1995, ApJ, 440, 742}

\bibitem[Hawley {et~al.}(1996)Hawley, Gammie, \& Balbus]{HGB96}
{---------. 1996, ApJ, 464, 690}

\bibitem[Kato, Fukue, \& Mineshige(1998)]{KFM98}
{Kato, S., Fukue, J., \& Mineshige, S. 1998, Black-Hole Accretion Disks.
Kyoto University Press (Kyoto)}

\bibitem[Kato \& Yoshizawa(1995)]{KY95}
{Kato, S. \& Yoshizawa, A. 1995, Publ. Astron. Soc. Jap., 47, 629}

\bibitem[King, Pringle, \& Livio(2007)]{KPL07}
{King, A.\ R., Pringle, J.\ E., \& Livio, M. 2007, MNRAS, 376, 1740}

\bibitem[Miller \& Stone(2000)]{MS00}
{Miller, K.\ A. \& Stone, J.\ M. 2000, ApJ, 534, 398}	

\bibitem[Ogilvie(2003)]{Ogilvie03}
{Ogilvie, G.\ I. 2003, MNRAS, 340, 969}

\bibitem[Pessah, Chan, \& Psaltis(2006a)]{PCP06a}
{Pessah, M.\ E., Chan C.\ K., \& Psaltis, D. 2006a, MNRAS, 372, 183}

\bibitem[Pessah, Chan, \& Psaltis(2006b)]{PCP06b}
{---------. 2006b, Phys. Rev. Lett., 97, 221103}

\bibitem[Pessah, Chan, \& Psaltis(2007)]{PCP07}
{---------. 2007, MNRAS, submitted, [astro-ph/0612404]}

\bibitem[Sano, Inutsuka, \& Miyama(1998)]{Sanoetal98}
{Sano, T., Inutsuka, S.\ I., \& Miyama, S.\ M. 1998, ApJ, 506, L57}

\bibitem[Sano {et~al.}(2004)Sano, Inutsuka, Turner, \& Stone]{Sanoetal04} 
{Sano, T., Inutsuka, S.\ I., Turner, N.\ J., \& Stone, J.\ M. 2004, ApJ, 605, 321}

\bibitem[Shakura \& Sunyaev(1973)]{SS73}
{Shakura, N.\ I. \& Sunyaev, R.\ A. 1973, A\&A, 24, 337}

\bibitem[Tan \& Blackman(2004)]{TB04}
{Tan, J.\ C. \& Blackman, E.\ G. 2004, ApJ, 603, 401}

\bibitem[Tout \& Pringle(1992)]{TP92}
{Tout, C.\ A. \& Pringle, J.\ E. 1992, MNRAS, 259, 604}


\end{thebibliography}
\end{document}